\documentstyle[12pt]{article}

\input{epsf}
 
\def\beq{\begin{equation}}
\def\eeq{\end{equation}}
\def\barr{\begin{eqnarray}}
\def\earr{\end{eqnarray}}

\def\bq{\begin{quote}}
\def\eq{\end{quote}}
\def\spose#1{\hbox to 0pt{#1\hss}}
\def\lsim{\mathrel{\spose{\lower 3pt\hbox{$\mathchar''218$}}
 \raise 2.0pt\hbox{$\mathchar''13C$}}}
\def\gsim{\mathrel{\spose{\lower 3pt\hbox{$\mathchar''218$}}
 \raise 2.0pt\hbox{$\mathchar''13E$}}}

\begin{document}

\begin{titlepage}
 
\begin{flushright}
CERN-TH/98-123\\
FERMILAB-PUB-98/101-T\\
IC/98/34\\
hep-ph/9804254
\end{flushright}

\vspace{0.15truecm}
\begin{center}
\boldmath
\large\bf
Resolving a Discrete Ambiguity in the CKM Angle $\beta$ through\\ 
\vspace{0.2truecm}
$B_{u,d}\to J/\psi\, K^\ast$ and $B_s\to J/\psi\, \phi$ Decays
\unboldmath
\end{center}
\vspace{0.3truecm}
\begin{center}
Amol S. Dighe\\[0.1cm]
{\sl The Abdus Salam International Centre for Theoretical Physics\\ 
34100 Trieste, Italy}\\[0.6cm]
Isard Dunietz\\[0.1cm]
{\sl Theoretical Physics Division, Fermi National Accelerator Laboratory\\
Batavia, IL 60510, USA}\\[0.6cm]
Robert Fleischer\\[0.1cm]
{\sl Theory Division, CERN, CH-1211 Geneva 23, Switzerland}
\end{center}

\vspace{0.15cm}

\begin{abstract}
\vspace{0.2cm}\noindent
It is well known that $\sin(2\beta)$, where $\beta$ is one of the angles of
the unitarity triangle of the CKM matrix, can be determined in a 
theoretically clean way by measuring mixing-induced CP violation in the decay 
$B_d\to J/\psi\,K_{\rm S}$. Another clean extraction of this CKM angle is 
provided by the time-dependent angular distribution for the decay products of 
$B_d\to J/\psi(\to l^+l^-)\,K^{\ast0}(\to \pi^0 K_{\rm S})$, where we have 
more observables at our disposal than in the case of $B_d\to 
J/\psi\,K_{\rm S}$, so that in addition to $\sin(2\beta)$ also $\cos(2\beta)$
can be probed in a direct way. Unfortunately a sign ambiguity remains in 
$\cos(2\beta)$. If it could be resolved, a discrete ambiguity in the 
extraction of the CKM angle $\beta$ could be resolved as well, which would
allow a more incisive test of the CKM model of CP violation. This note shows 
that detailed time-dependent studies of $B_{u,d}\to J/\psi\,K^{\ast}$ and 
$B_s\to J/\psi\,\phi$ decay processes can determine the sign of 
$\cos(2\beta)$, thereby removing the corresponding ambiguity in the 
extraction of the CKM angle $\beta$.
\end{abstract}

\vfill
\noindent
CERN-TH/98-123\\
April 1998
 
\end{titlepage}
 
\thispagestyle{empty}
\vbox{}
\newpage
 
\setcounter{page}{1} 

\noindent
The conventional methods for determining the angles $\alpha$, $\beta$ and
$\gamma$ of the usual unitarity triangle \cite{ut} of the 
Cabibbo--Kobayashi--Maskawa matrix (CKM matrix) \cite{ckm} leave 
several discrete ambiguities~\cite{fleischerreview}. This is also 
the case for the ``gold-plated'' mode $B_d\to J/\psi\,K_{\rm S}$. 
The mixing-induced CP asymmetry arising in this channel allows only a 
theoretically clean determination of $\sin(2\beta)$, so that a discrete 
four-fold ambiguity for the extracted value of $\beta \in [0^\circ,
360^\circ]$ remains. In the recent literature, several strategies were 
proposed to resolve ambiguities of this kind~\cite{ambig}.

Another clean probe of the CKM angle $\beta$ is provided by the observables 
of the angular distributions for the decay products of 
$B_d\to J/\psi(\to l^+l^-)\,K^{\ast0}(\to \pi^0 K_{\rm S})$ modes 
\cite{kayser}--\cite{ddf1}. Such observables can in general be expressed 
in terms of decay amplitudes as 
\begin{equation}\label{obs1}
|A_f(t)|^2,\quad\mbox{Re\,}[\,A_{\widetilde f}^\ast(t)\, A_f(t)\,], \quad
\mbox{Im\,}[\,A_{\widetilde f}^\ast(t)\, A_f(t)\,], 
\end{equation}
where $f$ and $\widetilde f$ are labels for specific final-state 
configurations. The full three-angle distributions for tagged 
$B_d(t)\to J/\psi\,K^{\ast0}\,(\to \pi^0 K_{\rm S})$ decays are given 
in~\cite{dqstl}--\cite{ddlr}. Throughout this note, by tagging we mean making
the distinction of initially, i.e.\ at $t=0$, present unmixed $B^0_{d,s}$ and 
$\overline{B^0_{d,s}}$ mesons. Weighting functions have been derived to 
extract the corresponding observables in an efficient way from experimental 
data~\cite{dqstl,ddf1}. The time evolution of the interference terms 
in (\ref{obs1}), i.e.\ of the real and imaginary parts of bilinear 
combinations of certain decay amplitudes, allows the
determination~\cite{ddf1} of $\sin(\delta_{1,2})$ and 
$\cos(\delta_1-\delta_2)$, 
where $\delta_1$ and $\delta_2$ are CP-conserving strong phases, and of 
$\sin(2\beta)$ and 
\begin{equation}\label{cos-obs}
\cos(\delta_{1,2})\cos(2\beta).
\end{equation}
The CP-conserving observables $|A_f(t = 0)|$, $\sin(\delta_{1,2})$ 
and $\cos(\delta_1-\delta_2)$ can be determined to a higher accuracy 
from the much larger data samples arising for $B^\pm\to J/\psi\,K^{\ast\pm}$ 
transitions, and untagged $B_d$ decays into $J/\psi\,K^{\ast0} 
(\to K^+ \pi^-)$ and $J/\psi\overline{K^{\ast0}} 
(\to K^- \pi^+)$ states \cite{dqstl,ddf1}. At first sight, one may think
that $\sin(\delta_{1,2})$ and $\cos(\delta_1-\delta_2)$ extracted this
way from the $B_{u,d}\to J/\psi\,K^{\ast}$ angular distributions will allow 
the determination of $\cos(2\beta)$ with the help of the terms given in 
(\ref{cos-obs}). A closer look shows, however, that this is unfortunately 
not the case, since we do not have sufficient information to fix the 
{\it signs} of $\cos(\delta_{1,2})$, thereby leaving a sign ambiguity for 
$\cos(2\beta)$.

The purpose of this letter is to point out that this ambiguity can
be resolved with the help of tagged, time-dependent studies of  
$B_s\to J/\psi\,\phi$ decays. The angular distributions are given 
in~\cite{ddf1, ddlr}, and weighting functions to extract the observables 
from experimental data can be found in~\cite{ddf1}. An important feature 
of these observables is that they allow the determination of a CP-violating 
weak phase $\phi$~\cite{ddf1, dsnowmass93}, which takes a very small value,
of ${\cal O}(0.03)$, within the Standard Model, and represents a sensitive 
probe for new-physics contributions to $B_s^0$--$\overline{B_s^0}$ 
mixing. Provided there is a sizeable mass difference between the mass
eigenstates $B_s^H$ and $B_s^L$, this phase can even be extracted from 
{\it untagged} $B_s$ data samples \cite{fd1}, where the rapid $\Delta m_s t$ 
oscillations cancel \cite{bsbsbar}. 

Another important feature is the fact that the tagged, time-dependent
$\stackrel{(-)}{B_s}(t)\to J/\psi(\to l^+l^-)\,\phi(\to K^+ K^-)$ 
observables corresponding to the ``Im'' terms in (\ref{obs1}) provide 
sufficient information to determine $\widehat\delta_{1,2}$ 
{\it unambiguously}, where the strong phases $\widehat\delta_{1,2}$ 
are the flavour $SU(3)$ counterparts of $\delta_{1,2}$. In the strict 
$SU(3)$ limit, we have $\widehat\delta_{1,2}=\delta_{1,2}$. The time 
evolution of these observables takes the following form~\cite{ddf1}:
\begin{equation}\label{obs2}
e^{-\overline{\Gamma}t} 
\sin(\widehat\delta_k - \Delta m_s t)\;\stackrel{(-)}{+} \,
\frac{1}{2}\left(e^{-\Gamma_H t}-e^{-\Gamma_L t}\right)
\cos(\widehat\delta_k)\,\phi\,,
\end{equation}
where terms of ${\cal O}(\phi^2)$ have been neglected, 
$\Gamma_H$, $\Gamma_L$ denote the decay widths of the $B_s$ mass
eigenstates, $\overline{\Gamma}\equiv (\Gamma_H+\Gamma_L)/2$, and $k=1,2$.
Consequently, the strong phases $\widehat\delta_{1,2}$ can be determined 
{\it unambiguously} by resolving the rapid $\Delta m_s t$ oscillations in
(\ref{obs2}). Comparing the resulting values for $\sin(\widehat\delta_{1,2})$ 
and $\cos(\widehat\delta_{1}-\widehat\delta_{2})$ with their unhatted 
analogues, which can be determined from the $B_{u,d}\to J/\psi\, K^\ast$ 
observables, we obtain valuable information on $SU(3)$ breaking. In order 
to fix the sign of $\cos(2\beta)$ with the help of (\ref{cos-obs}), we just 
need the sign of $\cos(\delta_1)$ or $\cos(\delta_2)$, which is provided by 
the sign of $\cos(\widehat\delta_k)$ determined from the 
$B_s(t)\to J/\psi\,\phi$ observables.

The $SU(3)$ flavour symmetry should work reasonably well to determine this
sign, unless $|\sin(\delta_k)|$ is close to 1, implying $\delta_k$ close to
$90^\circ$ or $270^\circ$, where $\cos(\delta_k)$ flips its sign. However,
for such values of $\delta_k$, the $\cos(\delta_k)\cos(2\beta)$ terms 
(\ref{cos-obs}) appearing in the $B_d\to J/\psi(\to l^+l^-)\,K^{\ast0}(
\to \pi^0 K_{\rm S})$ angular distribution -- which are essential for our
strategy -- will anyway be highly suppressed, so that it is doubtful that the 
sign ambiguity can be resolved in this case. It is of course not yet clear 
whether future experiments will encounter such an unfortunate situation. 
Within the framework of ``factorization'', we have 
$\delta_{1,2}\in\{0^\circ,180^\circ\}$, i.e.\ $|\cos(\delta_{1,2})|=1$. 
Since $B_{u,d}\to J/\psi\, K^\ast$ decays and their $B_s$ counterpart 
$B_s\to J/\psi\,\phi$ are colour-suppressed modes, ``factorization'' is not 
expected to be a good approximation in this case. Consequently, the actual 
values of $\delta_{1,2}$ may deviate significantly from these trivial values. 

Angular distribution measurements for $B\to J/\psi\,K^{\ast}$ modes 
have already been reported~\cite{cleo}, and others may soon be 
made public~\cite{schmidtpappas}.  The modes considered here 
are very appealing, because of the ability to trigger on the $J/\psi$ meson.  
The prospects are bright for resolving the rapid $\Delta m_s t$ oscillations
in $B_s(t)\to J/\psi\,\phi$ decays at planned experiments at the Tevatron 
and the LHC. Thus, in a not too distant future, the determination of 
$\widehat\delta_k$ and the resolution of the ambiguity (related to the 
sign of $\cos(2\beta)$) in the extraction of the CKM angle $\beta$ from 
$B_d\to J/\psi\,K^{\ast0}(\to\pi^0 K_{\rm S})$ decays may become feasible. 
Let us note that there remains a two-fold ambiguity for $\beta$ in this 
approach, since we cannot decide whether $\beta$ lies within the intervals 
$[0^\circ,180^\circ]$ or $[180^\circ,360^\circ]$. In each interval, 
$\beta$ is, however, fixed unambiguously. Consequently, the original 
four-fold ambiguity arising in the extraction of $\beta$ from $\sin(2\beta)$ 
can be reduced to just a two-fold ambiguity. Usually it is argued that 
$\varepsilon_K$, which measures indirect CP violation in the kaon system, 
implies the former range~\cite{ambig}.

While the $B_d\to J/\psi\,K^{\ast0}(\to\pi^0 K_{\rm S})$ mode is very 
accessible at $B$ factories operating at the $\Upsilon(4S)$ resonance, 
detectors at hadron accelerators should study the feasibility of the $\pi^0$ 
reconstruction. The $B_d\to J/\psi\,\rho^0$, $J/\psi\,\omega$ modes could 
be added to $B_d\to J/\psi\,K^{\ast0}(\to\pi^0 K_{\rm S})$ in order to 
resolve the $\beta$ ambiguity. If penguin amplitudes are neglected, 
the time evolution of these decay modes also depends on the CKM angle 
$\beta$ and, in the limit of the $SU(3)$ flavour symmetry, their strong 
phases are equal to those of their $SU(3)$ counterparts.  

In summary, traditional methods allow tests of the CKM picture of CP 
violation only up to discrete ambiguities. The resolution of these ambiguities 
would make such CKM tests significantly more powerful. In this letter, 
making use of the many observables that are available from angular 
correlations, we have proposed an approach to resolve a discrete ambiguity 
in the determination of the CKM angle $\beta$ that may be simpler than 
strategies advocated earlier~\cite{ambig}. More generally, 
angular-correlation methods can also be formulated to remove discrete CKM 
ambiguities in $\beta$, $2\beta+\gamma=\pi+\beta-\alpha$ and $\gamma$ 
from colour-allowed processes \cite{colourallowed}.

\vspace{0.4cm}

\noindent
This work was supported in part by the Department of Energy, 
Contract No.~DE-AC02-76CHO3000.

\end{document}